\documentclass[conference]{IEEEtran}
\IEEEoverridecommandlockouts
\usepackage{cite}
\usepackage{amsmath,amssymb,amsfonts}
\usepackage{algorithmic}
\usepackage{graphicx}
\usepackage{textcomp}
\usepackage{xcolor}
\usepackage{todonotes}
\usepackage{alltt}
\usepackage{listings}
\usepackage{hyperref}
\def\BibTeX{{\rm B\kern-.05em{\sc i\kern-.025em b}\kern-.08em
    T\kern-.1667em\lower.7ex\hbox{E}\kern-.125emX}}
\begin{document}

\title{Towards Turing-Complete Quantum Computing Coming From Classical Assembler
}

\author{\IEEEauthorblockN{Thomas Gabor, Marian Lingsch Rosenfeld, Claudia Linnhoff-Popien}
\IEEEauthorblockA{\textit{LMU Munich} \\
thomas.gabor@ifi.lmu.de, M.Rosenfeld@campus.lmu.de}
}

\maketitle

%\todo{Submission is double blind!}

\begin{abstract}
Instead of producing quantum languages that are fit for current quantum computers, we build a language from standard classical assembler and augment it with quantum capabilities so that quantum algorithms become a subset of it. This paves the way for the development of hybrid algorithms directly from classical software, which is not feasible on today's hardware but might inspire future quantum programmers.
\end{abstract}


\section{Introduction}

The capabilities of current quantum computers remain very limited, most importantly with respect to the depth of the executed algorithm. As astonishing as it is that we can still run notably important algorithms (maybe even better than classical ones~\cite{bravyi2018quantum,arute2019quantum}) on this kind of hardware, the algorithms that can be fully implemented on a quantum computer lack several characteristics that programmers are used to for classical algorithms, most notably: recursion. The absence of arbitrary-depth recursion (or simply jumps) prohibits pure quantum algorithm from emulating every algorithm that can be run on a Turing machine.\footnote{We leave out the discussion that finite-memory systems can never execute all algorithms a Turing machine can execute.} For today's quantum computers, this is a more than sensible assumption as decoherence times only allow for a couple of gates to be executed in sequence before all useful information is lost. But what if we had the best quantum computer we could wish for?

In contrast to most other approaches (most notably quantum assembly languages~\cite{cross2021openqasm}) we attempt to design an assembly language not for quantum computers with today's capabilities but for an imaginary quantum computer that is not limited by today's challenges in hardware design and maybe not even limited by all the rules of quantum mechanics at all times. However, we argue that such a language (even though conceivably it will not run on any real quantum computer for a long time) can still be useful for two main reasons:

\begin{itemize}
	\item First, for now and at least for the immediate future a large amount of quantum algorithms will be run on classical simulators. As these classical simulators run on classical machines, it is reasonable to write algorithms that can access these classical machines' capabilities directly and thus forego any quantum simulation when possible. This may produce algorithms that -- while not compatible with (current) quantum computers -- may at least not be as superfluously slow when run on a quantum simulator for testing purposes or simply due to a lack of better quantum options.
	\item Second, we need tools to discover new quantum algorithms that are more accessible to programmers well versed in standard programming languages. As of now, most setups of quantum software are built around a few algorithm archetypes which are known or suspected to provide some form of quantum advantage. As these few archetypes already hold up all the fascination regarding quantum computers, the discovery of new archetypes should be a big step towards more usage scenarios for quantum computers. Enabling programmers to write an algorithm classically first and then translate it part by part to quantum-suitable reformulations (while maintaining a runnable program) could enable new creativity in algorithm design. 
\end{itemize}

\section{Approach}

In this work-in-progress paper we provide a very simple assembly language that we augmented with a few quantum-specific instructions and capabilities. Table~\ref{tab:instructions} shows all instructions at our disposal. We provide basic mathematical and logical operations as well as the ``set'' and ``swap'' instructions. We then provide a set of instructions to manipulate program flow including the very simple ``setpc'' and ``jump'' instructions and the more elaborate ``ifte'' instruction, which provides the functionality of a standard if-then-else expression in most programming languages. However, all of our registers can be quantum registers, which means that they can contain a superposition instead of a single scalar value. This also means that our program counter can be in superposition, which causes the entire program state to be in superposition and possibly execute different instructions in each of its superposed branches.

To actually introduce said superpositions into the registers, Table~\ref{tab:instructions} also shows a few quantum-specific instructions. Most notably, we can apply a Hadamard gate to a single qubit or a range of qubits. We also provide a shortcut for the diffusion operator used within Grover's algorithm as well as for phase multiplication.

Obviously, this set of instruction is far from complete even for classical algorithms but it allows us to formulate a few simple examples of how to use our language.\footnote{See \url{https://github.com/marian-lingsch/quantum-assembler}.}

\begin{table}[t]
\centering
  \begin{tabular}{p{9em} p{19em}}
  	\hline
      Classic Instructions & \\ \hline
      add d$i$ d$j$ d$k$ & add the value of cell $i$ with the value of cell $j$ and write the result to cell $k$ \\
      mul d$i$ d$j$ d$k$ & multiply the value of cell $i$ with the value of cell $j$ and write the result to cell $k$ \\
      div d$i$ d$j$ d$k$ & divide the value of cell $i$ with the value of cell $j$ and write the result to cell $k$ \\
      sub d$i$ d$j$ d$k$ & substract the value of cell $j$ from the value of cell $i$ and write the result to cell $k$ \\
      sqrt d$i$ d$j$ & write the square root of the value of cell $i$ into cell $j$ \\
      mod d$i$ d$j$ d$k$ & writes into cell $k$ the modulo of the value of cell $i$ with respect to the value of cell $j$ \\

      neg d$i$ & negate the boolean value of the cell $i$ \\
      and d$i$ d$j$ d$k$ & write into cell $k$ the boolean value resulting from v$i$ $\land$ v$j$, where v$l$ is the value of cell d$l$ \\
      or d$i$ d$j$ d$k$ & write into cell $k$ the boolean value resulting from v$i$ $\lor$ v$j$, where v$l$ is the value of cell d$l$ \\

      set d$i$ $j$ & set the value of cell $i$ to $j$ \\
      swap d$i$ d$j$ & swap the values of cells $i$ and $j$ \\
      setpc d$i$ & set the program counter to the value of cell $i$ \\
      jump $i$ &  set the program counter to the value of $i$ \\
      skip & do nothing \\
      stop & remain in this instruction indefinitely \\
      ifte d$i$ $j$ $k$ & if cell $i$ is true, set the pc to the value of $j$ else set the pc to the value of $k$ \\ \hline

      Quantum Instructions & \\ \hline
      havoc d$i$ $j$ $k$ & apply Hadamard gates to the qubits in range from $j$ to $k$ of cell $i$ \\
      havocb $i$ & apply a Hadamard gate to the data qubit $i$ \\
      diffusion & apply Grover's diffusion operator \\
      phase $x$ $y$ & multiply the phase of the state by $x + \mathrm{i}\mkern1mu \cdot y$ \\
  \end{tabular}
  \vspace{1em}
  \caption{Instruction set for assembly with quantum capabilities}
  \label{tab:instructions}
  \vspace{-1em}
\end{table}

\section{Examples}

Listing~1 shows a simple example of what we can easily produce when we apply superposition to the program counter. Lines 12--15 all multiply the value in d4 by 2 (see line~6). However, we do not always execute all of them. Instead, we initialize the program counter to the start of these lines (see line~5) and then add the equally distributed superposition of the values $\langle 0,1,2,3 \rangle$ to that value. When all superposed branches of the program have been executed, we end up with an equal superposition of values $\langle 1, 2, 4, 8 \rangle$ saved in register d4. We were able to define such a superposed value using patterns familiar from imperative programming and the approach can be generalized to more complicated superposed program flows.

Listing~2 shows how we could evolve quantum algorithms from classical ones. We use a staple example of quantum computing, i.e., finding the prime factors of a given number; but we start out with the most common classical approach to that problem, i.e., the sieve of Eratosthenes~\cite{hoare1972proof}. If we assume that in the classical world we would simply derive the prime factors of a number by trying out all possible divisors, it appears evident that such a search for a divisor might be accelerated by employing Grover's algorithm~\cite{mandviwalla2018implementing}. Naturally, this is a bit counterintuitive to anyone familiar with Shor's algorithm~\cite{lanyon2007experimental}, which we will discuss in a bit. However, the approach still works: In Listing~2, we assume that the number we want to split into factors is given as ``NUMBER1'' (see line~4) and just to keep things simple we assume that said number's bit count is given as ``NUMBER2''. Obviously, a slightly more powerful assembly language could just derive that information with a single instruction. Line~8 then creates a superposition of all numbers of the same length as our input ``NUMBER1'' and with line~9 we ensure that this superposition no longer contains the values $0$ and $1$ so that we can try all remaining values as possible divisors. In line~10 we again utilize the superposed value in d1 to compute the remainder for all divisors. We then set up the parameters for Grover's search (lines 11--16) and subsequently perform the calculated amount of iterations within Grover's search, looking for a single $0$ remainder within the superposed value of d2. Note that we implemented the instructions ``phase'' and ``diffusion'' within the instruction set specifically to enable a straightforward definition of algorithms like Listing~2 based on Grover's search. Specialized instructions like these not only increase readability but might also allow future versions of the simulator to implement classical shortcuts for certain behavior.

\lstdefinelanguage{tcqc} {morekeywords={set,havoc,add,mod,sqrt,div,mul,ifte,phase,diffusion,skip,stop,setpc,sub,jump}, sensitive=false,
morecomment=[l]{;}}

\lstset{language=tcqc}
\lstset{frameround=tttt}

\begin{lstlisting}[float,caption={Nondeterministic multiplication},basicstyle=\small,captionpos=b,frame=single,numbers=left, numberstyle=\tiny, stepnumber=2, numbersep=5pt,belowcaptionskip=2pt]
; nondeterministic multiplication
; produces multiple powers of 2
start:
    havoc d0 0 2
    set d1 subMul
    set d2 2
    set d4 1
    add d0 d1 d3
    setpc d3

subMul:
    mul d4 d2 d4
    mul d4 d2 d4
    mul d4 d2 d4
    mul d4 d2 d4
    stop
\end{lstlisting}

\begin{lstlisting}[float,caption={Factoring via Grover},basicstyle=\small,captionpos=b,frame=single,numbers=left, numberstyle=\tiny, stepnumber=2, numbersep=5pt,belowcaptionskip=2pt]
start:
    ; set number for which is to
    ;   be factorized
    set d0 NUMBER1
    ; havoc is used in range format
    ; each int is represented by 8 bits
    ; NUMBER2 equals log2(sqrt(NUMBER1))
    havoc d1 0 NUMBER2
    add d1 2 d1
    mod d0 d1 d2
    ; d3 is the amount of iterations
    ;   of Grover's algorithm
    add d0 0 d3
    sqrt d3 d3
    div d3 4 d3
    mul d3 3 d3

grover:
    ifte d2 jumpTrue jumpFalse

jumpFalse:
    phase -1.0 0.0
    jump subDiffusion

jumpTrue:
    skip
    jump subDiffusion

subDiffusion:
    diffusion
    sub d3 1 d3
    ifte d3 subStop grover

subStop:
    stop
\end{lstlisting}

We compared the quasi-quantum program in Listing~2 to its purely classical counterpart, for which we do not provide an additional listing to save space (and because it looks entirely as expected). It is important to note that we also wrote the classical algorithm for the sieve of Eratosthenes using the same quantum-enabled assembly language and ran it using the same quantum simulator \cite{gabor2022simple}. Figure~1 shows the respective run times of both algorithms for the whole range of inputs. We can see that the classical algorithm takes the longest on actual prime numbers where it needs to try out every possible divisor while the quantum algorithm needs to produce the superposition of all possible divisors for every single input value regardless. However, it is interesting to note that within our simulator, both approaches show a remarkably similar performance. This justifies our approach to offer a ``soft transition'' from a purely classical execution of a purely classical algorithm to a classical simulation of an algorithm with quantum parts without necessarily blow up the run time disproportionally.

\begin{figure}[t]
\centering
\includegraphics[width=.95\columnwidth]{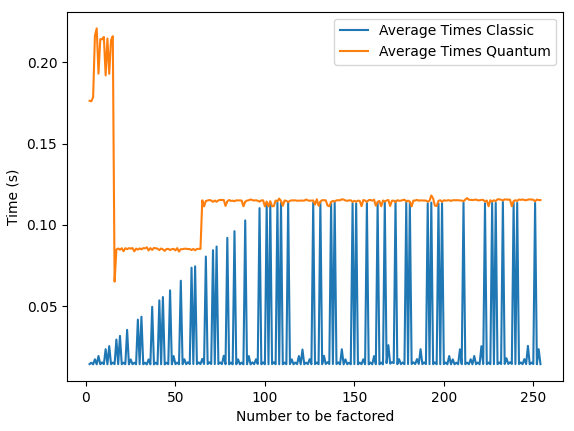}
\caption{Comparison between classical and quantum factoring algorithms written in our assembly language.}
\label{fig:benchmark}
\end{figure}

\section{Conclusion}

The assembly language shown here is very limited but is intended to provide an idea how to re-think hybrid programming coming from full-fledged classical languages. We have analyzed how quantum program flow might look like on an assembly level and how it may allow to construct complex quantum states in an intuitive and imperative manner (Listing~1). We have also presented how very versatile quantum approaches like Grover's search can be integrated into well-known classical algorithms in a straightforward way (Listing~2), even though we sacrifice the ability to run the resulting algorithms on today's quantum computers immediately. In that case, we can also see where the approach still falls short: While implementing Grover's search here may provide some benefits, quantum programmers know that Shor's algorithm provides an even better way to solve the factorization problem on a quantum computer. However, that algorithm is not as easily derived from its classical counterpart -- or at least our language does not yet allow to do so. Future versions of the language should provide more versatile and more powerful instruction sets in order to provide a ``soft transition'' into quantum software for more quantum algorithms.

In principle, there is no reason to keep this ``transition language'' at an assembly level; in fact, for usability it might be especially important to translate our concepts of quantum program flow to more high-level language models (and perhaps offer automatic translation back to assembly level to maintain compatibility). An ideal case might be to allow a programmer to use a modern programming language and insert quantum-compatible expressions one by one until an interpreter or compiler notices that a certain module is now fully quantum-compatible and then offers the option to also run it on quantum hardware (compare Tyagi et al.~\cite{tyagi2016toward} for the case of reversible computing). However, high-level abstractions for quantum computing that are both intuitive and powerful are still very much sought after.

%\section*{Acknowledgment}
%
%The preferred spelling of the word ``acknowledgment'' in America is without 
%an ``e'' after the ``g''. Avoid the stilted expression ``one of us (R. B. 
%G.) thanks $\ldots$''. Instead, try ``R. B. G. thanks$\ldots$''. Put sponsor 
%acknowledgments in the unnumbered footnote on the first page.


\end{document}